\documentclass{article}

\usepackage{arxiv}
\usepackage[T1]{fontenc}
\usepackage[utf8]{inputenc}
\usepackage{textcomp}
\usepackage{amsmath,amssymb}
\usepackage{booktabs}
\usepackage{graphicx}
\usepackage{microtype}
\usepackage{cite}
\usepackage[hidelinks]{hyperref}

\title{DED-Agent: An agentic AI method for autonomous thermal simulation model development and calibration in laser directed energy deposition}
\renewcommand{\shorttitle}{DED-Agent for L-DED thermal simulation}
\renewcommand{\headeright}{arXiv preprint}

\author{%
Michael Song\textsuperscript{1} \qquad Yilong Dai\textsuperscript{1} \qquad Lin Cheng\textsuperscript{*}\\[4pt]
{\normalfont Department of Mechanical Engineering, University of Maryland, College Park, MD 20742, USA}\\[2pt]
{\normalfont\normalsize \textsuperscript{1}These authors contributed equally to this work.}\\
{\normalfont\normalsize \textsuperscript{*}Corresponding author. E-mail: \texttt{lcheng90@umd.edu}}
}

\date{November 23, 2026}

\begin{document}
\maketitle

\begin{abstract}
In metal additive manufacturing (AM), process parameters set the local thermal history, which governs defects, microstructure and properties. Because laser power, scan speed, hatch spacing, layer thickness, scan strategy and position combine into a design space of 10\textsuperscript{4}--10\textsuperscript{6} conditions, mapping it with finite element (FE) thermal models is beyond what human experts can do by hand: every model must be set up, meshed, solved and calibrated by an expert, and calibration of unknown heat-source parameters alone can take days to weeks. This paper presents DED-Agent, an agentic artificial intelligence (AI) method for autonomous simulation model development and calibration for laser directed energy deposition (L-DED) process. In this multi-agent system, a supervisor agent coordinates three specialist agents that share a structured simulation specification and call open-source tools to achieve reliable simulation. Specifically, an intent agent turns a plain-language request into a validated specification, a simulation agent builds, runs and checks an FE thermal model through physics-aware step-by-step reasoning, and a calibration agent identifies the key parameters of heat-source models against experimental data (e.g., thermocouple data) using a high-order proper generalized decomposition (HOPGD) surrogate within a trust-region search. On five benchmark builds, from single-track walls to a multi-track band on a cylindrical tube, the agents built executable thermal models without manual intervention, with 4.1--9.1\% error for the simple and medium builds. Within three calibration rounds the error fell to 0.9--2.5\% for these builds. The surrogate matched full FE to within 0.08--1.55\% and shortened calibration by 5--13 times. These results show that agentic AI can autonomously establish reliable and accurate simulation models for AM processes, and has great potential to accelerate the exploration and identification of optimal process parameters for microstructure control.
\end{abstract}


\section{Introduction}

In metal additive manufacturing (AM), processing conditions determine the local thermal history, which in turn governs microstructure evolution, defect formation and, ultimately, the material properties and performance of the part \cite{r1,r2}. Because the thermal history varies from track to track, from layer to layer and from one location to another, a single build can exhibit columnar, mixed and equiaxed grains and a spatial variation of mechanical properties. A quantitative understanding of how processing conditions shape the thermal history, and how the thermal history in turn controls the microstructure, is therefore essential for process optimization and for ensuring the reliability of AM components \cite{r2,r3}.

Over the past two decades, thermal simulation models have been developed to understand the thermal history of AM and how it connects to microstructure and properties \cite{r1,r2}. The first part-scale models borrowed heavily from welding. They represented the laser as a moving volumetric heat source, typically the Goldak double ellipsoid \cite{r4}, and solved transient heat conduction with temperature-dependent properties and latent heat. What sets AM apart from welding is the continuous addition of material, which these models handle by activating elements along the toolpath. Michaleris \cite{r5} showed that the choice between quiet and inactive elements, together with the treatment of the evolving free surface, changes the predicted temperatures and heat losses. Once validated against in-situ thermocouple measurements, these finite element (FE) models reproduced the layer-by-layer heating and heat accumulation in directed energy deposition (DED) walls, and coupling them with mechanical analyses made it possible to predict distortion and residual stress \cite{r3,r6}. At the smaller scale of the melt pool, heat and fluid flow models capture Marangoni convection and the resulting melt-pool geometry \cite{r7}, while mechanistic models link the computed temperature gradient and solidification rate to grain morphology, defects and properties \cite{r2}. The cost of these high-fidelity models, however, limits their use in process design. Recent work has therefore developed reduced-order and data-driven surrogates and embedded them in digital twins \cite{r8}, including proper generalized decomposition approaches that construct parametric solutions from a few high-fidelity runs \cite{r9,r10,r11}. With these faster models, thermal simulation is increasingly used to interpret in-situ monitoring data, select process parameters and scan strategies, and control interlayer temperature and cooling rate \cite{r1,r8}.

The ability of metal AM to fabricate complex geometries, together with its large number of adjustable process parameters, offers great potential for tailoring the local microstructure and properties within a single part. However, the same flexibility makes it difficult to develop reliable thermal simulation models for every build of interest. Laser power, scan speed, hatch spacing, layer thickness, scan strategy and spatial position combine into a high-dimensional process space; each combination produces a different thermal history in space and time, and each thermal history can lead to a different microstructure. The resulting design space, $N_{proc}\times N_{spatial}\times N_{mechanism}$, easily exceeds $10^{4}$--$10^{6}$ conditions. Physics-based FE thermal models are the main tool for exploring it \cite{r3,r6,r8}, but each model is still built case by case by an expert. The analyst sets up the geometry and toolpath, applies boundary conditions, meshes the domain, configures the solver and post-processes the results. A mistake at any step either crashes the solver or, worse, produces a model that runs but silently represents the wrong physics. The heat-source model \cite{r4} also contains parameters, such as absorption efficiency that depend on melt-pool size and are unknown before a build is printed and characterized \cite{r3,r12}. These parameters are calibrated manually or with global optimizers that require hundreds of FE runs, so a single calibrated model can take days to weeks. At this cost, human experts can study only a handful of points in a design space of millions, and much of the potential of AM for site-specific property control remains unexplored.

Agentic artificial intelligence (AI) offers a promising route to overcome this bottleneck. Agents built on large language models (LLMs) can reason about a task, invoke external tools and iteratively act on the results \cite{r13,r14}, and they have begun to automate scientific computing workflows, including multi-agent systems that solve mechanics problems \cite{r15} and agents that set up and run computational fluid dynamics simulations in OpenFOAM \cite{r16,r17,r18}. However, these systems largely automate the downstream workflow once the mathematical model has been specified, whereas the more fundamental step of translating a physical problem into a correct numerical model remains challenging. Directly prompting an LLM to write the solver is not yet reliable: in our recent benchmark of 75 partial differential equation (PDE) problems, frontier LLMs achieved scores of about 88/100 on physical reasoning, yet none of the FEniCSx solvers they generated executed successfully end to end \cite{r19}. To the best of our knowledge, no agentic method for metal AM has yet closed the loop from a user's intent to an executable thermal model and its calibration against experimental measurements.

To address this gap, this paper presents DED-Agent, a multi-agent AI method for the autonomous development and calibration of thermal simulation models in laser directed energy deposition (L-DED). Each agent in DED-Agent has a clearly defined role, input and output. Specifically, a supervisor agent coordinates three specialist agents: (i) an intent agent that turns a plain-language request into a validated simulation specification, (ii) a simulation agent that builds, runs and checks an FE thermal model through physics-aware, step-by-step reasoning, and (iii) a calibration agent that identifies heat-source parameters from thermocouple data with a high-order proper generalized decomposition (HOPGD) surrogate \cite{r10,r11}. We first describe the design of each agent and then evaluate the method on five benchmark L-DED builds of increasing geometric and toolpath complexity. The present work focuses on autonomous model development and calibration against published experimental data, whereas calibration against in-house experiments, agent-based cooling-rate control and defect minimization will be reported in a subsequent journal paper.

\section{DED-Agent method}

\subsection{Supervisor agent and shared state}

Figure 1 presents the DED-Agent, which consists of a supervisor agent and specialist agents. The supervisor turns the user's request into a task plan, assigns each subtask to a specialist, retries failed steps and escalates to the user when information is missing or a check fails repeatedly. The specialists do not pass free text to one another. They read and write a shared state that holds the simulation specification (JSON), run records, measurements and logs. Every hand-off passes automatic checks: schema validation, physics sanity checks (units, property ranges, energy input, peak temperature relative to the liquidus), self-correction with a retry limit and, if necessary, escalation to the user. The agents call a tool layer of open-source components: a temperature-dependent material database ($k$, $ c_{p}$, $\rho$, latent heat, solidus and liquidus), Gmsh \cite{r20} for meshing, FEniCSx/DOLFINx \cite{r21} with PETSc for the FE solution, post-processing of probe histories, melt pool and cooling rate, the HOPGD surrogate, and differential evolution \cite{r22} and Nelder--Mead \cite{r23} optimizers. All agents in DED-Agent use Grok 4.6 as the LLM backbone, and all experiments ran on one workstation: AMD Ryzen 9 9900X CPU (12 cores, 24 threads), 32 GB RAM and an NVIDIA GeForce RTX 5070 Ti GPU (16 GB, CUDA 13.0, driver 595.99). The finite-element solves ran on the GPU in a single process through cuDOLFINx, using GMRES with an algebraic multigrid (GAMG) preconditioner and diagonal scaling.

\subsection{Intent agent}

The intent agent is the interface between the user and the simulation. It reads a natural-language request such as ``run a new thermal simulation and plot the temperature history at one probe'' followed by a description of the base plate, wall and toolpath, together with a CAD model of the part to be printed, and extracts the material, substrate and deposit geometry, toolpath, laser power, scan speed, layer height, dwell time, boundary conditions and probe locations. Values missing from the request are filled from the material database or documented defaults and flagged as assumptions, and all values are checked for units and physical ranges. The output is a JSON specification, for example \{``material'': ``SS316L'', ``laser\_power\_W'': 800, ``scan\_speed\_mm\_s'': 16.7, ``layers'': 20, ``probes'': [``TC2'']\}, which is the single source of truth for all downstream agents. If a quantity is ambiguous and cannot be defaulted safely, for example the location of a thermocouple, the supervisor asks the user before any simulation is run.

\begin{figure}[ht]
\centering
\includegraphics[width=0.98\textwidth]{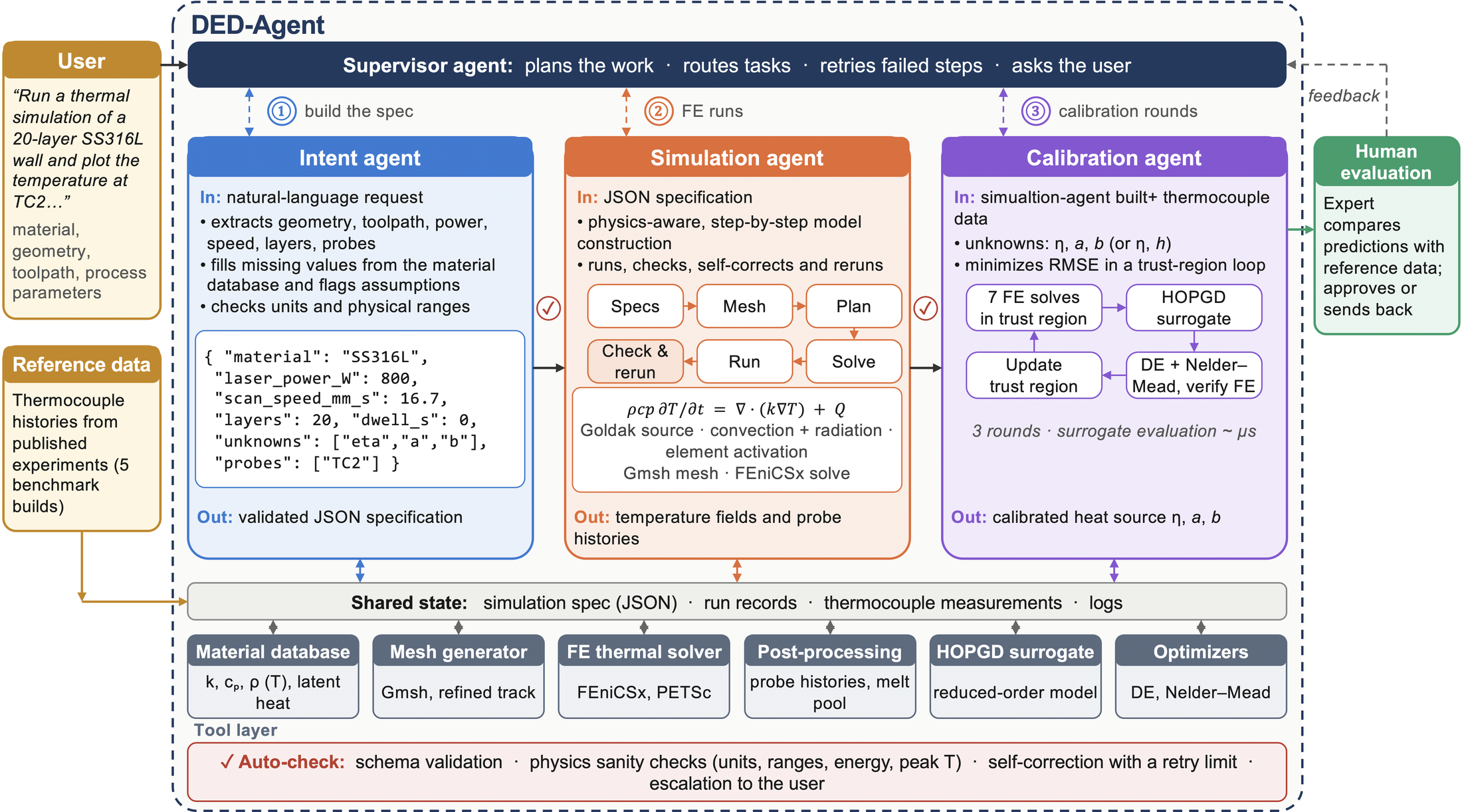}
\caption{Architecture of DED-Agent. The supervisor agent plans the work, routes tasks, retries failed steps and asks the user when needed. The intent agent (\textcircled{1}) turns the request into a validated JSON specification; the simulation agent (\textcircled{2}) builds, runs and checks the FE model by step-by-step physics-to-code reasoning; the calibration agent (\textcircled{3}) identifies the heat-source parameters in a trust-region loop with a HOPGD surrogate. The agents share a common state, call the tool layer and pass automatic checks at every hand-off; a human expert evaluates the result. Cooling-rate control, defect minimization and calibration with in-house experiments are future work.}
\label{fig:architecture}
\end{figure}

\subsection{Simulation agent}

The simulation agent turns the specification into an executable, checked FE model. Rather than generating solver code end to end, it follows the step-by-step reasoning we developed for PDE-to-FEM solver synthesis \cite{r19}. It parses the specification, assembles the governing equation, locates and types each boundary and initial condition, selects the discretization and solution strategy, and defines the post-processing quantities. It then writes the code following a modular plan: geometry, mesh import with tags, function space, variational form, constraints, solution and post-processing.

For L-DED, the agent constructs the transient heat conduction problem for the temperature $T\left(x,t\right)$ on the active domain $\Omega \left(t\right)$:

\begin{equation}
\rho c_{p}\frac{\partial T}{\partial t}=\nabla \cdot \left(k\nabla T\right)+Q\left(x,t\right), \quad \text{in } \Omega \left(t\right)
\label{eq:heat}
\end{equation}

where $\rho$, $ c_{p}$ and $ k$ are the temperature-dependent density, specific heat and conductivity, with latent heat represented by an apparent heat capacity between the solidus and liquidus. The laser is modeled by a Goldak ellipsoidal volumetric source \cite{r4}:

\begin{equation}
Q=\frac{6\sqrt{3}\eta P}{abc\pi \sqrt{\pi }}\exp \left(-\frac{3\xi ^{2}}{c^{2}}-\frac{3y^{2}}{a^{2}}-\frac{3z^{2}}{b^{2}}\right)
\label{eq:goldak}
\end{equation}

where $P$ is the laser power, $\eta$ the absorption efficiency, $ a$ the transverse semi-axis, $ b$ the depth semi-axis, $ c$ the longitudinal semi-axis, $\xi$ and $ y$ the in-plane coordinates relative to the moving source ($\xi$ along the scan direction), and $z$ the vertical coordinate measured from the top surface of the deposit ($ z \leq 0$ below it). Free surfaces lose heat by convection and radiation:

\begin{equation}
-k\nabla T\cdot n=h\left(T-T_{\infty }\right)+\varepsilon \sigma \left(T^{4}-T_{\infty }^{4}\right), \quad \text{on } \Gamma _{s}\left(t\right)
\label{eq:bc}
\end{equation}

where $h$ is the heat transfer coefficient, $\varepsilon$ the emissivity and $\sigma$ the Stefan--Boltzmann constant. Deposition is modeled by element activation along the toolpath (inactive-element method) \cite{r5}, with the free surface $\Gamma _{s}\left(t\right)$ updated as elements are activated. Deposit elements are meshed from the start, but before activation their conductivity and heat capacity are scaled by factors of 10\textsuperscript{-6} and 10\textsuperscript{-2}, and the source and surface losses are switched off on them. An element is activated when its centroid is at or below the current layer top and the normalized Goldak intensity at any of its nodes reaches 0.05. The agent meshes the substrate and deposit in Gmsh with refinement along the track, discretizes Eq. (1) with first-order Lagrange elements and backward-Euler time integration in FEniCSx, and selects the time step so that the source advances a quarter of the transverse semi-axis per step. After each run it checks solver convergence, non-physical temperatures and energy balance. If a check fails, it diagnoses the cause, edits the code and reruns within the retry limit. It returns probe temperature histories, temperature fields and melt-pool metrics.

\subsection{Calibration agent}

Even when the agent-built model is physically correct, its predictions depend on heat-source parameters that cannot be measured before printing, and these parameters are typically the dominant source of error. The calibration agent therefore adjusts them so that the simulated thermal history agrees with in-situ measurements. Global optimizers such as genetic algorithms are commonly used for this purpose, but they require hundreds of FE evaluations \cite{r12}, which is impractical for part-scale L-DED models. The agent instead calibrates the key heat-source parameters, $\mu = (\eta, a, b)$, with the longitudinal semi-axis set equal to the transverse one ($ c = a$), or the absorption efficiency together with boundary-condition parameters when the source geometry is already known. To keep the number of FE solves small, the agent replaces the FE model with a reduced-order surrogate. Following Lu et al. \cite{r10,r11}, the surrogate is built with high-order proper generalized decomposition (HOPGD), a variant of proper generalized decomposition \cite{r9}, which constructs a non-intrusive, separated representation of the probe temperature from a small number of FE snapshots:

\begin{equation}
T\left(t;\mu \right)\approx \sum_{m=1}^{M} F_{m}^{t}\left(t\right)\prod_{j=1}^{d} F_{m}^{j}\left(\mu _{j}\right)
\label{eq:hopgd}
\end{equation}

where $F_{m}$ are one-dimensional modes, $ M$ is the number of modes and $ d = 3$ is the number of parameters. Once constructed, the surrogate can be evaluated in microseconds anywhere within its parameter range, which makes an exhaustive optimization affordable. Calibration is then posed as minimizing the root-mean-square error (RMSE) between the simulated and measured thermocouple histories within a trust-region box $ B_{k}$ in round $ k$:

\begin{equation}
\min _{\mu \in B_{k}} J\left(\mu \right)=\sqrt{\frac{1}{N}\sum_{i=1}^{N} \left(T\left(t_{i};\mu \right)-T_{i}^{\exp }\right)^{2}}
\label{eq:rmse}
\end{equation}

Each calibration round proceeds as follows. The agent first runs FE simulations on a sparse stencil in $B_{k}$ and uses these snapshots to build the HOPGD surrogate. Eq. (5) is then minimized on the surrogate, with differential evolution providing a global search and Nelder--Mead refining the result. Because the surrogate is only an approximation, the resulting optimum is verified with a full FE solve. Finally, the agent re-centers and resizes $ B_{k}$ around the best verified point. Three rounds are performed in total. Accuracy is also reported as the mean relative (percent) error. To assess how strongly the measurements constrain each parameter, the normalized sensitivity of the RMSE at the identified point is computed as

\begin{equation}
S_{j}=\left|\frac{\partial J}{\partial \mu _{j}}\right|\Delta \mu _{j}
\label{eq:sens}
\end{equation}

where $\Delta \mu _{j}$ is the width of the admissible range of $\mu _{j}$.

\section{Benchmark cases}

To evaluate DED-Agent, five L-DED builds with published thermocouple data were assembled into a case database and grouped into three levels of geometric and toolpath complexity. Table 1 lists the five cases together with their sources, complexity levels, build descriptions and calibrated parameters. The simple level comprises two single-track walls (s1, s3); the medium level comprises a multi-track square deposit (m1) and a single wall built without interlayer dwell (m2); and the complex level is a multi-track band deposited by a zigzag scan on a cylindrical tube (c2). For each case, the agents received only a plain-language description of the build, with no template simulation model, and their predictions were compared with the measured and published temperature histories by a human expert.

\begin{table}[ht]
\centering
\caption{Benchmark cases.}
\label{tab:benchmark}
\small
\setlength{\tabcolsep}{4pt}
\resizebox{\textwidth}{!}{%
\begin{tabular}{lllll}
\toprule
ID & Source & Level & Build & Calibrated parameters \\
\midrule
s1 & Yang et al. \cite{r6} & Simple & Ti-6Al-4V single-track rectangular wall, 10 layers & $\eta$, convective boundary \\
s3 & Nain et al. \cite{r24} & Simple & SS316L single-track wall (60 mm) on 100 $\times$ 50 mm plate & $\eta$, a, b \\
m1 & d'Ambelle et al. \cite{r25} & Medium & SS316L multi-track square deposit, 20 layers, zigzag, no dwell, 800 W, 1 m/min, 0.38 mm/layer & $\eta$, a, b \\
m2 & Heigel et al. \cite{r3} & Medium & Ti-6Al-4V single wall, no dwell (case 1) & $\eta$, a, b \\
c2 & Ye et al. \cite{r26} & Complex & SS316L multi-track band on a cylindrical tube, zigzag scan & $\eta$, a, b \\
\bottomrule
\end{tabular}}
\end{table}

\section{Results and discussion}

The results are presented in the order of the agent chain. Section 4.1 examines how the intent agent translates user requests into simulation specifications, Section 4.2 evaluates the thermal models built and executed by the simulation agent, and Section 4.3 assesses how the calibration agent corrects these models against measurements. Each subsection first examines a representative case in detail and then summarizes the results for all five cases. The workflow underlying Sections 4.1 and 4.2 is illustrated in Fig. 2.

\subsection{Intent agent: from request to specification}

Figure 2 shows a typical request, here for case m1, which specifies the material, deposit geometry, number of layers, scan strategy, dwell time, laser power, scan speed, layer height and the probe at which the temperature is to be reported, together with the information of build part. From requests of this kind, the intent agent produced a complete JSON specification for all five cases, each of which passed schema validation before being handed to the simulation agent. The completeness of the request, however, has a direct effect on simulation accuracy. For case s3, a fully specified prompt yielded an RMSE of 51.8 \textdegree{}C and a 9.13\% error relative to the experiment, as shown in the s3 row of Fig. 3, whereas an under-specified prompt, for which the agent had to assume default values for several process parameters, yielded 74.0 \textdegree{}C (13.12\%). Because the agent flags every assumed value, the user can identify the inputs that most affect the prediction, while the remaining uncertainty in the heat source is subsequently addressed by the calibration agent.

\begin{figure}[ht]
\centering
\includegraphics[width=\textwidth]{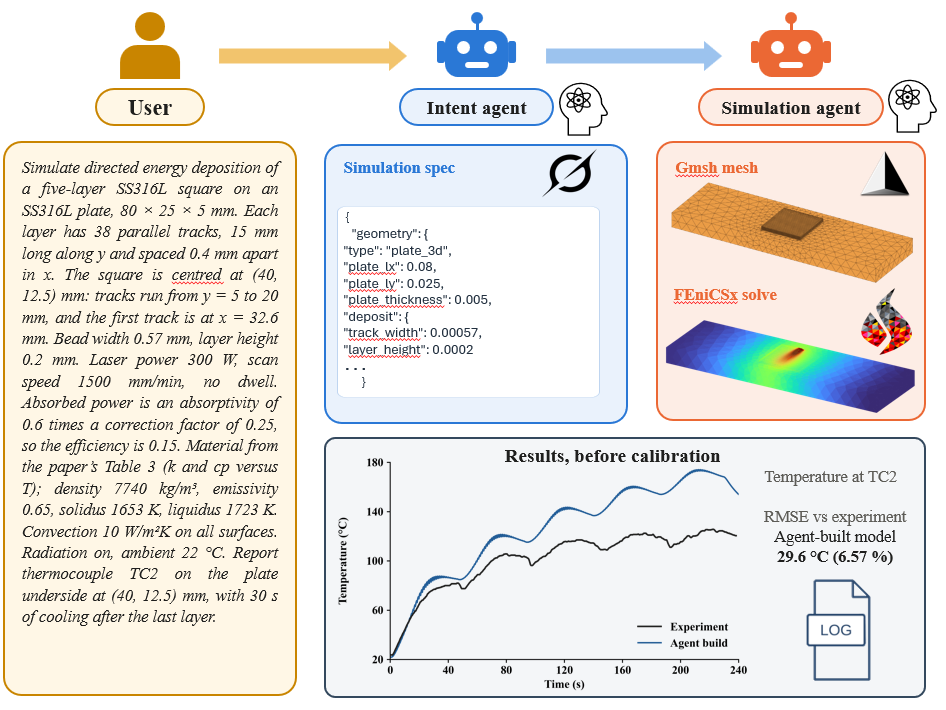}
\caption{Autonomous model development, illustrated for case m1 (d'Ambelle et al. \cite{r25}, SS316L multi-track wall, 5 layers, probe 2). From left to right: the user submits a plain-language request with no template simulation model; the intent agent converts the request into a JSON specification that is checked against the schema, units and admissible ranges, with default values filled in and flagged; the simulation agent meshes the domain in Gmsh, solves the transient thermal problem in FEniCSx, and checks every run automatically, rerunning it on failure; and a human expert compares the predicted and measured temperature histories at probe 2, giving an RMSE of 29.6 \textdegree{}C and a relative error of 6.57\%. Log file stores the token usage and computer wall-time.}
\label{fig:workflow}
\end{figure}

\subsection{Simulation agent: autonomous model development}

To illustrate how the simulation agent operates, a case of medium-complexity multi-track square deposit (m1), through the simulation stage of the workflow was shown in Fig. 2. From the validated specification provided by intent agent, simulation agent built the substrate and the 5-layer multi-track square deposit, generated a mesh in Gmsh with refinement along the deposition path, and assembled the transient thermal model of Eqs. (1) - (3) in FEniCSx, with elements activated along the rectangular contour toolpath. Before returning any result, the agent checked the solver convergence, the temperature range and the energy balance of the run, and reran the model automatically whenever a check failed. The final model ran in 46.2 min without manual intervention. The predicted temperature history at TC2 reproduces the layer-wise heating and cooling cycles and the gradual heat accumulation measured in the experiment, but over-predicts the interlayer temperature, with an RMSE of 29.6 \textdegree{}C (6.57\%). This discrepancy is expected when nominal process parameters are used, and it is addressed by the calibration agent in Section 4.3.

The same procedure was then applied to all five benchmark cases. Fig. 2 shows the result for m1, and Fig. 3 compares, for each of the remaining four cases, the reference geometry and measured data with the mesh, temperature field and probe history generated by the simulation agent, and Table 2 summarizes the corresponding model statistics. All five models were built and executed without manual intervention. For the simple cases, the models reproduced both the layer-by-layer heating peaks and the subsequent cooling. For s1, the predicted history at probe 1 is comparable to that of the FE model in the original study, with an RMSE of 15.4 \textdegree{}C (4.05\%), and for s3 the RMSE was 51.8 \textdegree{}C (9.13\%) for the fully specified prompt. The other medium case, m2, showed the same trend as m1: the model captured the gradual heat accumulation and the layer wise temperature oscillations but over-predicted the interlayer temperature (62.0 \textdegree{}C, 8.40\%). The complex case c2 was the most demanding, requiring the agent to construct a cylindrical substrate, a curved multi-track band and a toolpath with repeated zigzag reversals. The predicted history at point E reproduces the timing and sequence of the thermal peaks but under-predicts the peak magnitudes and the interpass temperature (RMSE 362.8 K, 20.9\%). As summarized in Fig. 3, the pre-calibration error was 4.05--9.13\% for the four simple and medium builds and 20.9\% for c2.

Table 2 reports, for each case, the model size, the number of self-correction cycles and the types of errors that the agent detected and corrected before obtaining a valid run. The most common error was an excessively large time step: in m1 and c2, the initial step let the source advance too far per increment, which the validator rejected and the agent corrected. In c2, a unit error in the conductivity table was also detected and corrected.

\begin{table}[ht]
\centering
\caption{Simulation-agent statistics.}
\label{tab:simulation}
\small
\setlength{\tabcolsep}{4pt}
\resizebox{\textwidth}{!}{%
\begin{tabular}{lllllll}
\toprule
Case & Elements & Time steps & Self-corrections & Error types fixed & Token usage (Input/Output) & Wall time \\
\midrule
s1 & 118,850 & 8,144 & 0 & -- & 3282/1291 & 1.4 min \\
s3 & 132,464 & 4,959 & 0 & -- & 3030/1089 & 8.9 min \\
m1 & 72,243 & 40,781 & 1 & Time step & 8906/2718 & 46.2 min \\
m2 & 345,543 & 6,891 & 0 & -- &  & 19.9 min \\
c2 & 3,433,793 & 1,410 & 3 & Time step, unit &  & 5.5 min \\
\bottomrule
\end{tabular}}
\end{table}

\begin{figure}[ht]
\centering
\includegraphics[width=\textwidth]{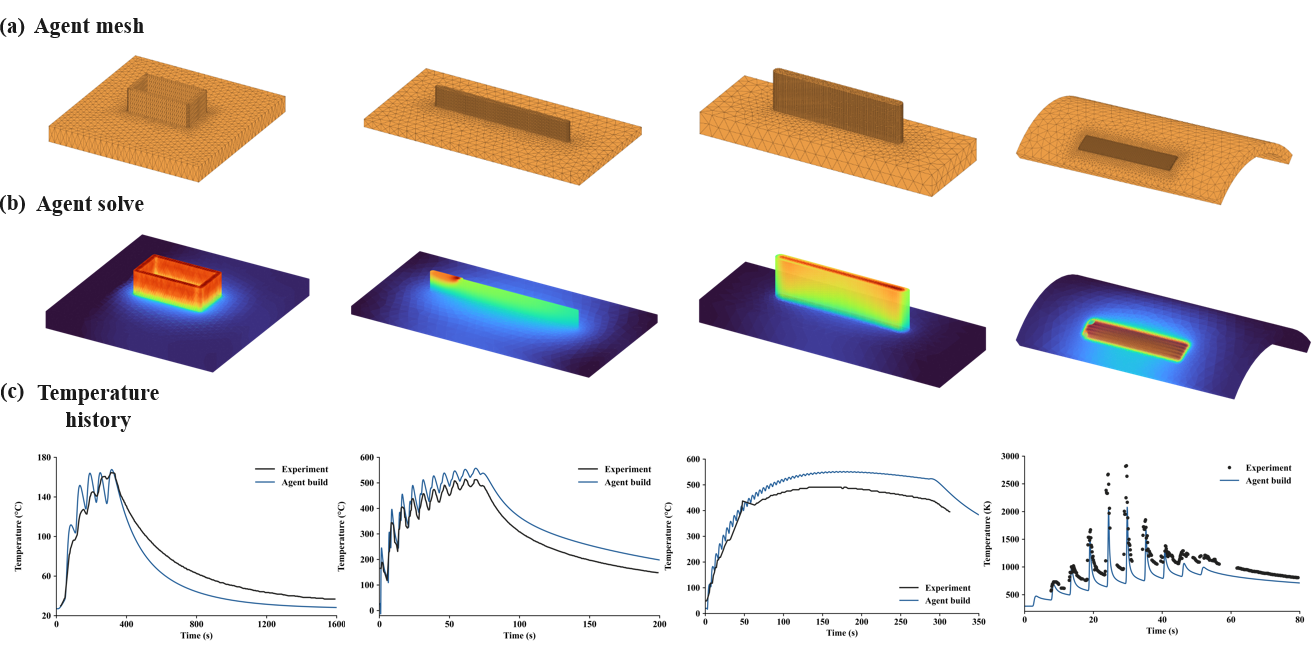}
\caption{Models developed autonomously by the simulation agent for the remaining four benchmark cases (from left to right: s1, s3, m2 and c2; case m1 is shown in Fig. 2): (a) agent-generated mesh; (b) agent-predicted temperature field; (c) agent-predicted temperature history compared with the reference.}
\label{fig:models}
\end{figure}

\subsection{Calibration agent: autonomous calibration}

To illustrate the calibration procedure, we consider m1 as a specific case, whose uncalibrated model over-predicted relatively the interlayer temperature in Section 4.2. Figure 4 illustrates the complete chain for this case. User gives general prompt to describe the overview and objective of problem. The intent agent first extracted the build setup (a 5-layer multi-track square deposit on an SS316L plate, unidirectional scan with no dwell, P300W, S25mm/s, 0.2 mm per layer) and the problem statement (absorption efficiency and the transverse and depth semi-axes unknown; compute the RMSE against the experiment). From this specification, the simulation agent built the initial model (round 0), which over-predicted the heat accumulation at thermocouple TC2 (RMSE 29.6 \textdegree{}C, 6.57\%). The calibration agent then built star plan and sampled the trust region, constructed the HOPGD surrogate and updated the heat-source parameters on it, by following repeat theses steps in future rounds until complete the last round. The error decreased to 4.86\% after round 1 and to 1.93\% after round 2, and the round-3 optimum, verified by a full FE solve, reached 7.88 \textdegree{}C (1.87\%). The calibrated history closely follows both the measured layer-wise oscillations and the gradual rise in interlayer temperature.

\begin{figure}[ht]
\centering
\includegraphics[width=\textwidth]{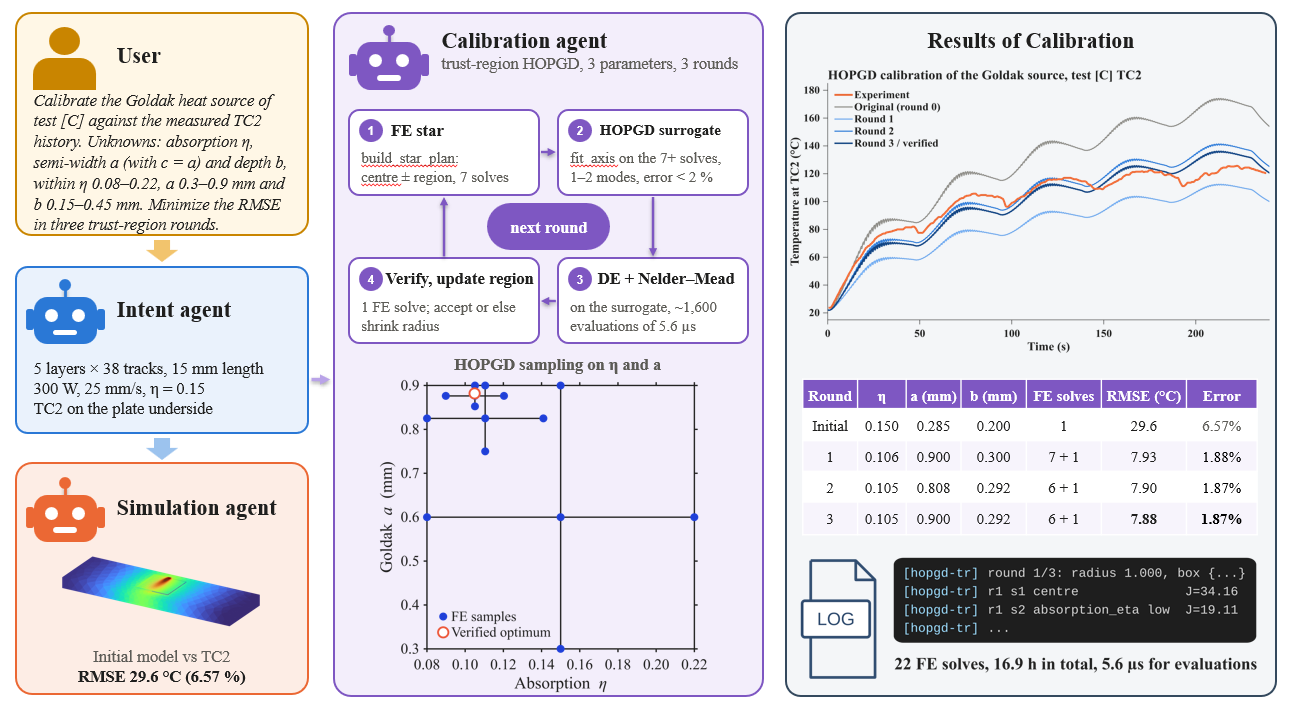}
\caption{Agent chain for case m1. Agent 1 (intent) states the setup and problem; agent 2 (simulation) builds the initial model; agent 3 (calibration) samples the trust region and builds the HOPGD surrogate. Right: TC2 histories of the experiment, the original model and rounds 1--3, with the RMSE and percent error versus round.}
\label{fig:chain}
\end{figure}

The same calibration was then applied to all five cases, and Fig. 5 and Table 3 summarize the results. For case s1, the heat-source dimensions had already been calibrated in the original study, so the agent calibrated the absorption efficiency and the convective boundary condition instead. In every case, the largest improvement occurred in the first round, and the subsequent rounds converged. For the four simple and medium builds, calibration reduced the RMSE by 64.5--73.4\% to 5.46--19.92 \textdegree{}C, corresponding to percent errors of 0.89--2.48\%. For c2, calibration reduced the percent error from 20.90\% to 8.26\%, but the RMSE remained high at 271 \textdegree{}C. Because this residual persisted even as the source parameters were varied over a wide range, it is attributed to model-form limitation, namely the treatment of the zigzag reversals, heat dissipation and boundary conditions on the curved geometry, which cannot be corrected by calibrating the heat source alone.

\begin{figure}[ht]
\centering
\includegraphics[width=\textwidth]{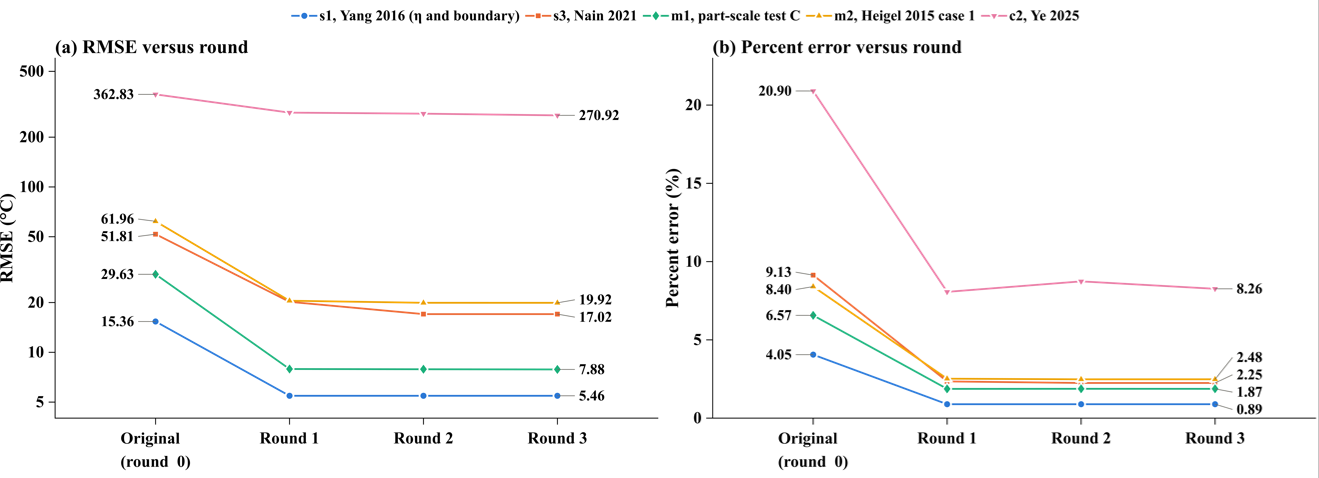}
\caption{Convergence of the trust-region HOPGD calibration for the five cases: (a) RMSE (log scale) and (b) percent error versus round.}
\label{fig:convergence}
\end{figure}

\begin{table}[ht]
\centering
\caption{Calibration performance summary.}
\label{tab:calibration}
\small
\setlength{\tabcolsep}{4pt}
\resizebox{\textwidth}{!}{%
\begin{tabular}{llllll}
\toprule
Case & RMSE & Error (\%) & Identified $\eta$, a (mm), b (mm) & Surrogate vs. FE history (\%) & FE / HOPGD time (h) \\
\midrule
s1 & 15.4 $\rightarrow$ 5.5 \textdegree{}C & 4.05 $\rightarrow$ 0.89 & $\eta$ and h (0.4,5) & 4.06 & 2.3 / 0.5 ($\times$5) \\
s3 & 51.8 $\rightarrow$ 17.0 \textdegree{}C & 9.13 $\rightarrow$ 2.25 & 0.40, 1.35, 0.30 & 0.19 & 24.0 / 3.3 ($\times$7) \\
m1 & 29.6 $\rightarrow$ 7.9 \textdegree{}C & 6.57 $\rightarrow$ 1.87 & 0.105, 0.90, 0.292 & 0.37 & 162 / 16.9 ($\times$10) \\
m2 & 62.0 $\rightarrow$ 19.9 \textdegree{}C & 8.40 $\rightarrow$ 2.48 & 0.397, 2.00, 1.80 & 1.19 & 96.0 / 7.3 ($\times$13) \\
c2 & 362.8 $\rightarrow$ 270.9 K & 20.90 $\rightarrow$ 8.26 & 0.65, 2.39, 0.625 & 1.65 & 10.6 / 2.0 ($\times$5) \\
\bottomrule
\end{tabular}}
\end{table}

Because each calibration round relies on the HOPGD surrogate rather than on the FE model, the accuracy and computational benefit of the surrogate were also evaluated. At every verification point, the surrogate predictions were compared with full FE solutions. The RMSE predicted by HOPGD tracked with the FE RMSE over two orders of magnitude, and in the final round the surrogate reproduced the FE temperature histories to within 0.19-4.06\%. Case s1 shows higher difference with FE history, due to agent-build model are more accurate towards ground truth than FE model. A single HOPGD evaluation required 6--8 \textmu{}s, compared with 1.4--46.2 min (cases of different complexity) for a full FE solve, corresponding to a speed-up of 1.1 $\times$ 10\textsuperscript{7} to 4.6 $\times$ 10\textsuperscript{8}. The end-to-end saving is smaller, because the surrogate must be rebuilt from new FE snapshots in each round. Even so, as listed in Table 3, a three-round calibration required 0.5--16.9 h, 5--13 times less than the estimated cost of a Nelder--Mead search on the full FE model. The savings were largest for the most computationally expensive models (m1, m2), which suggests that the advantage will grow with model size.

\subsection{Outlook: from model development to process exploration}

With DED-Agent, the development and calibration of a thermal model, which is the step that limits how much of the process design space can be explored, runs largely without manual intervention, leaving the human expert to review the result. Three extensions are under way. First, the models will be calibrated against in-house L-DED experiments that combine in-situ thermocouple and infrared measurements with ex-situ melt-pool cross-sections, which will also constrain the penetration depth of the source. Second, the optimization agent shown in Fig. 1 will search laser power, scan speed and dwell time on the calibrated surrogate to achieve a target cooling rate, and the recommended parameters will be verified by full FE simulation and by experiment. Third, cooling-rate control will be coupled with porosity and microstructure models to minimize defects. In addition, the benchmark will be extended to a wider range of geometries, materials and toolpaths, with repeated runs to quantify the variability of the agents' outputs, while model-form diagnosis and Bayesian calibration with a discrepancy term \cite{r27} will be used to address cases such as c2.

\section{Conclusions}

We presented DED-Agent, an agentic AI method in which a supervisor coordinates intent, simulation and calibration agents to turn a plain-language description of an L-DED build into an executable, calibrated FE thermal model. Across five benchmark builds, the intent agent converted plain-language requests into validated specifications, and the completeness of the request measurably affected accuracy, with errors of 9.1\% and 13.1\% for the fully and under-specified s3 prompts, respectively. Using physics-aware, step-by-step reasoning and execution checks, the simulation agent built executable thermal models for all cases without manual intervention, with pre-calibration errors of 4.05--9.13\% for the simple and medium builds. The HOPGD-based calibration agent then reduced these errors to 0.89--2.48\% within three rounds, while matching full FE to within 0.08--1.55\% and shortening calibration by a factor of 5--13. The residual 8.3\% error of the complex case c2 reveals a model-form limit, which motivates model-refinement, cooling-rate control and defect-minimization agents, validated against in-house experiments, in future work.

\section*{Acknowledgements}

The authors gratefully acknowledge financial support from the National Science Foundation (NSF, Award No. 2245106) and computational resources provided by the University of Maryland's Division of Information Technology.

\end{document}